\documentclass[
  aps,
  prd,
  twocolumn,
  superscriptaddress,
  nofootinbib,
  longbibliography,
  floatfix
]{revtex4-2}

\usepackage{amsmath,amssymb}
\usepackage{mathtools}
\usepackage{physics}
\usepackage{slashed}
\usepackage{bm}
\usepackage{graphicx}
\usepackage{mathrsfs}
\usepackage{braket}
\usepackage{cancel}
\usepackage{xcolor}
\usepackage{tikz}
\usepackage{etoolbox}
\usetikzlibrary{arrows.meta,decorations.pathreplacing,shapes.geometric,shapes.misc,calc}

\definecolor{colA}{HTML}{2F6690}   
\definecolor{colB}{HTML}{1B7F79}   
\definecolor{colC}{HTML}{B5442A}   
\definecolor{colGrey}{HTML}{4A4A4A}

\tikzset{
  vacuummode/.style   ={cross out, draw=colB, line width=0.85pt,
                     inner sep=0pt, minimum size=5.4pt},
  vacuumhole/.style   ={circle, draw=colB, fill=white, line width=0.9pt,
                     inner sep=0pt, minimum size=5.4pt},
  solhole/.style   ={circle, draw=colA, fill=white, line width=1.3pt,
                     inner sep=0pt, minimum size=6.8pt},
  pole/.style      ={circle, draw=colC, fill=colC,
                     inner sep=0pt, minimum size=4.6pt},
  chain/.style     ={draw=colC!60, line width=0.8pt,
                     dash pattern=on 2pt off 2pt},
  guide/.style     ={draw=colGrey!45, line width=0.6pt,
                     dash pattern=on 1.6pt off 1.8pt},
  vacuumline/.style   ={draw=colB, line width=1.5pt},
  imagevacuum/.style  ={draw=colB!45, line width=1.2pt,
                     dash pattern=on 3pt off 2.5pt},
  realline/.style  ={draw=colA, line width=1.5pt},
  axs/.style       ={->, draw=colGrey, line width=0.8pt},
  tk/.style        ={draw=colGrey, line width=0.7pt},
  lbl/.style       ={font=\footnotesize, text=colGrey},
  lblA/.style      ={font=\footnotesize, text=colA},
  lblB/.style      ={font=\footnotesize, text=colB},
  lblC/.style      ={font=\footnotesize, text=colC},
  bkflow/.style    ={->, draw=colB!80, line width=0.8pt},
  brc/.style       ={decorate, decoration={brace, amplitude=3pt},
                     draw=colC, line width=0.8pt}
}

\pretocmd{\section}{\addvspace{0.7\baselineskip}}{}{}
\pretocmd{\subsection}{\addvspace{0.5\baselineskip}}{}{}
\pretocmd{\subsubsection}{\addvspace{0.4\baselineskip}}{}{}

\begin{document}

\title{On-shell renormalization of sine-Gordon by the quantum inverse scattering method}

\author{Francesco Beccarini}
\email{francesco.beccarini@uniroma1.it}
\author{Claudio Conti}
\affiliation{Dipartimento di Fisica, Sapienza Universit\`a di Roma,
  Piazzale Aldo Moro 5, 00185 Roma, Italy}

\date{\today}
\begin{abstract}
In the quantum inverse scattering method, the sine-Gordon model is solved on a lattice, and its continuum and infinite-volume limits become uniform only after the length of the box is rescaled. We show that this rescaling, supplemented by an on-shell condition, defines a renormalization scheme. The whole dependence on the cut-off is absorbed into a factor $Z_L$ multiplying the length of the box, with an exponent fixed by the scaling dimension of the vertex operator; neither the field nor the coupling $\beta$ is renormalized, no subtraction scale is introduced, and the mass parameter receives a finite renormalization, fixed by requiring that the lightest breather has the mass of the boson in the linearized theory. The soliton and breather masses follow from the eigenvalues of the monodromy operator as functions of parameters that do not run, and reduce to the classical and semiclassical results as $\beta\rightarrow0$. In particular, the mass ratios are produced directly from the regularized construction, while the relation between the lattice parameters and the physical scale is absorbed into $Z_L$ and never needs to be computed. The scheme is a reparameterization of the theory which makes explicit the interplay between integrability and renormalization. We compare it with Coleman's normal ordering and with conformal perturbation theory, and we argue that it extends to ultralocal integrable theories with a massive spectrum, possibly including asymptotically free models that admit an ultralocal lattice regularization.
\end{abstract}

\maketitle

\section{Introduction}
\label{sec:introduction}

Two-dimensional integrable field theories are the only relativistic field theories whose spectrum, scattering amplitudes and Hilbert space can be constructed exactly and non-perturbatively. Classically, the construction is the inverse scattering method (ISM) \cite{PhysRevLett.19.1095,Ablowitz:1991xb,Faddeev:1987ph}, which trades the non-linear field for scattering data with linear time evolution, that is, for action--angle variables. Its quantum counterpart, the quantum inverse scattering method (QISM) \cite{Sklyanin:1979pfu,PhysRevD.19.3660,Thacker:1980ei,Korepin:1993kvr}, promotes the monodromy matrix to an operator whose entries obey an algebra fixed by an $R$-matrix, and builds the eigenstates of the conserved charges as Bethe states.

For a relativistic theory such as the sine-Gordon (SG) model, the construction requires a lattice regularization \cite{Sklyanin:1979pfu}: the vacuum has to be built by filling negative-energy quasiparticles, and the asymptotic particles -- solitons, antisolitons and breathers -- are assembled out of them. In \cite{Sklyanin:1979pfu} it was observed that the continuum and infinite-volume limits of the lattice model are not independent, and that they become uniform only after the box length has been rescaled, at the price of a finite ambiguity in the mass parameter. This observation was not developed further. The renormalization of the SG model is usually formulated in other terms: by normal ordering the cosine with respect to a free field \cite{Coleman:1974bu}, or, in conformal perturbation theory, by normalizing the vertex operators conformally and relating the coupling to the physical mass through exact results from integrability \cite{Zamolodchikov:1989ASPM,ZamolodchikovAl1995,LukyanovZamolodchikov1997}.

In this work we show that the rescaling of \cite{Sklyanin:1979pfu}, supplemented by an on-shell condition, defines a renormalization scheme, and we work out its consequences. All the dependence on the lattice spacing goes into a single factor $Z_L$ multiplying the length of the box, with an exponent fixed by the scaling dimension of the perturbing vertex operator. As expected, this procedure leaves the mass parameter $m$ defined only up to a finite multiplicative constant, which we fix by requiring that the lightest breather have the mass of the boson in the linearized theory. The soliton and breather masses are then functions of $(m,\beta)$ computed within the same regularized construction: their ratios are eigenvalues of the monodromy operator, and only their overall scale is fixed by the on-shell condition. Two intervals of the coupling are involved: the theory extends to $0<\beta^2<2\pi$, whereas the on-shell condition, and hence the scheme, is defined for $0<\beta^2<\pi$, where the first breather belongs to the spectrum. The scheme does not change the physics of the model; its interest lies in the parametrization it provides, which unveils the interplay between renormalization and integrability.

In Sec.~\ref{sec:quantisation} we summarize what it means to quantize an integrable field theory, contrast the non-relativistic and relativistic cases, and collect the results of \cite{Sklyanin:1979pfu} that we need: the vacuum rapidity density and the eigenvalues of the monodromy operator on the asymptotic states. In Sec.~\ref{sec:renormalization} we discuss the rescaling of the box length, formulate the renormalization scheme, derive the masses, and compare the scheme with normal ordering, conformal perturbation theory and other lattice regularizations. Conclusions are drawn in Sec.~\ref{sec:conclusions}.

\section{Quantization of integrable field theories}
\label{sec:quantisation}

The quantization of an integrable theory is structurally the same for a system of finitely many bodies and for a field with linear or non-linear interactions: one passes to action--angle variables and promotes them to operators. In this section we review the scheme, summarized in Fig.~\ref{fig:quantisation-scheme}, identify the point at which a relativistic theory differs from a non-relativistic one, and collect the results on the SG model on which Sec.~\ref{sec:renormalization} is built.

\subsection{Integrability and the action--angle scheme}
\label{sec:action-angle-scheme}

Integrability at the classical level means that there exists a canonical transformation to action--angle variables in which each action is an integral of motion \cite{Faddeev:1987ph}. For a system with $N$ degrees of freedom, the Liouville--Arnold theorem makes the statement sharp: $N$ functionally independent charges in involution are enough, their level sets are tori, and on each torus the motion is a linear flow. A field theory has infinitely many degrees of freedom and needs infinitely many conserved charges. In the action--angle chart, the equations of motion are trivial,
\begin{equation}
  \dot I_i = -\frac{\partial H}{\partial \varphi_i} = 0,
  \qquad
  \dot \varphi_i = \frac{\partial H}{\partial I_i} = \omega_i(I),
  \label{eq:action-angle-hamilton}
\end{equation}
so that the non-trivial content of the dynamics is pushed entirely into the mapping from the field to the action--angle variables.

\begin{figure*}[t]
\centering
\resizebox{\textwidth}{!}{%
\begin{tikzpicture}[
  >=Stealth,
  font=\small,
  box/.style={rounded corners=5pt, draw=#1, line width=1pt, fill=#1!7,
              align=center, inner sep=9pt, text width=6.4cm},
  box/.default=colGrey,
  boxI/.style={box=#1, minimum height=2.1cm},
  boxII/.style={box=#1, minimum height=3.5cm},
  boxIII/.style={box=#1, minimum height=3.0cm},
  arr/.style={->, line width=1.2pt, draw=#1},
  arr/.default=colGrey,
  lab/.style={midway, fill=white, inner sep=3.5pt, font=\footnotesize\itshape, text=#1},
  lab/.default=colGrey,
  qmark/.style={midway, circle, draw=#1, line width=1pt, fill=white,
                inner sep=2pt, minimum size=0.85cm, font=\Large\bfseries, text=#1},
  rowlab/.style={anchor=east, font=\footnotesize\scshape, text=black!50, align=right},
  hd/.style={font=\bfseries}
]

\def\cA{0}      \def\cB{7.8}    \def\cC{15.6}
\def\rI{0}      \def\rII{-5.0}  \def\rIII{-10.8}

\node[rowlab] at (-4.1,\rI)   {classical\\ theory};
\node[rowlab] at (-4.1,\rII)  {action--angle\\ form};
\node[rowlab] at (-4.1,\rIII) {quantum\\ theory};

\node[boxI=colA] (A1) at (\cA,\rI)
  {{\bfseries Finite number of bodies}\\[3pt]
   {\footnotesize $N$ degrees of freedom}\\[1pt]
   {\footnotesize\itshape (Toda, Calogero--Moser, \dots)}};

\node[boxI=colB] (B1) at (\cB,\rI)
  {{\bfseries Linear field theories}\\[3pt]
   {\footnotesize $\infty$ degrees of freedom}\\[1pt]
   {\footnotesize\itshape (free scalar, Maxwell, \dots)}};

\node[boxI=colC] (C1) at (\cC,\rI)
  {{\bfseries Non-linear field theories}\\[3pt]
   {\footnotesize $\infty$ degrees of freedom}\\[1pt]
   {\footnotesize\itshape (NLS, SG, KdV, \dots)}};

\node[boxII=colA] (A2) at (\cA,\rII)
  {{\bfseries Action--angle variables}\\[6pt]
   $\{\,I_i,\varphi_i\,\}_{i=1}^{N}$\\[7pt]
   $\displaystyle \dot I_i=-\frac{\partial H}{\partial\varphi_i}=0$\\[7pt]
   $\displaystyle \dot\varphi_i=\frac{\partial H}{\partial I_i}=\omega_i(I)$};

\node[boxII=colB] (B2) at (\cB,\rII)
  {{\bfseries Action--angle variables}\\[6pt]
   $\{\,I_k,\varphi_k\,\}_{k=1}^{\infty}$\\[7pt]
   $\displaystyle \dot I_k=-\frac{\partial H}{\partial\varphi_k}=0$\\[7pt]
   $\displaystyle \dot\varphi_k=\frac{\partial H}{\partial I_k}=\omega_k$};

\node[boxII=colC] (C2) at (\cC,\rII)
  {{\bfseries Action--angle variables}\\[5pt]
   {\footnotesize scattering data $\{\,a(\lambda),\,b(\lambda)\,;\,\lambda_n,\,c_n\,\}$}\\[6pt]
   $\mathcal{I}(\lambda)=\tfrac{1}{\pi}\log|a(\lambda)|$\\[4pt]
   $\varphi(\lambda)=\arg b(\lambda)$\\[6pt]
   $\dot{\mathcal{I}}(\lambda)=0, \qquad \dot\varphi(\lambda)=\omega(\lambda)$};

\node[boxIII=colA] (A3) at (\cA,\rIII)
  {$\hat I_i \, \qquad a_i \, \qquad a_i^{\dagger}$\\[8pt]
   $[\,a_i,a_j^{\dagger}\,]=\delta_{ij}$\\[10pt]
   $\mathcal{H}_{\mathrm{Weyl}}$};

\node[boxIII=colB] (B3) at (\cB,\rIII)
  {$\hat I_k \, \qquad a_k \, \qquad a_k^{\dagger}$\\[8pt]
   $[\,a_k,a_{k'}^{\dagger}\,]=\delta_{kk'}$\\[10pt]
   $\displaystyle \mathcal{H}_{\mathrm{Fock}}=\bigoplus_{N=0}^{\infty}\mathcal{H}_N$};

\node[boxIII=colC] (C3) at (\cC,\rIII)
  {$\hat{\mathcal{I}}(\lambda) \, \quad Z(\lambda) \, \quad Z^{\dagger}(\lambda)$\\[8pt]
   {\footnotesize $Z(\lambda)Z^{\dagger}(\mu)=
     S(\lambda,\mu)\,Z^{\dagger}(\mu)Z(\lambda)+2\pi\delta(\lambda-\mu)$}\\[10pt]
   $\displaystyle \mathcal{H}_{\mathrm{ZF}}=\bigoplus_{N=0}^{\infty}\mathcal{H}_N
    \quad {\scriptstyle (\ast)}$};

\draw[arr=colA] (A1) -- node[lab=colA] {canonical transformation} (A2);
\draw[arr=colB] (B1) -- node[lab=colB] {Fourier transform} (B2);
\draw[arr=colC] (C1) -- node[qmark=colC] {?} (C2);

\draw[arr=colA] (A2) -- node[lab=colA] {quantization map} (A3);
\draw[arr=colB] (B2) -- node[lab=colB] {quantization map} (B3);
\draw[arr=colC] (C2) -- node[lab=colC] {quantization map} (C3);

\end{tikzpicture}}

\caption{Each column starts from a classical integrable theory (top level), passes through its action--angle
description (middle level), and ends in the corresponding quantum theory (bottom level). The first arrow
is the canonical transformation that trivializes the dynamics; the second is the quantization map, which
is the same in all three columns. The diagram is detailed in the text; the asterisk is discussed in
Sec.~\ref{sec:number-operator}.
}
\label{fig:quantisation-scheme}
\end{figure*}
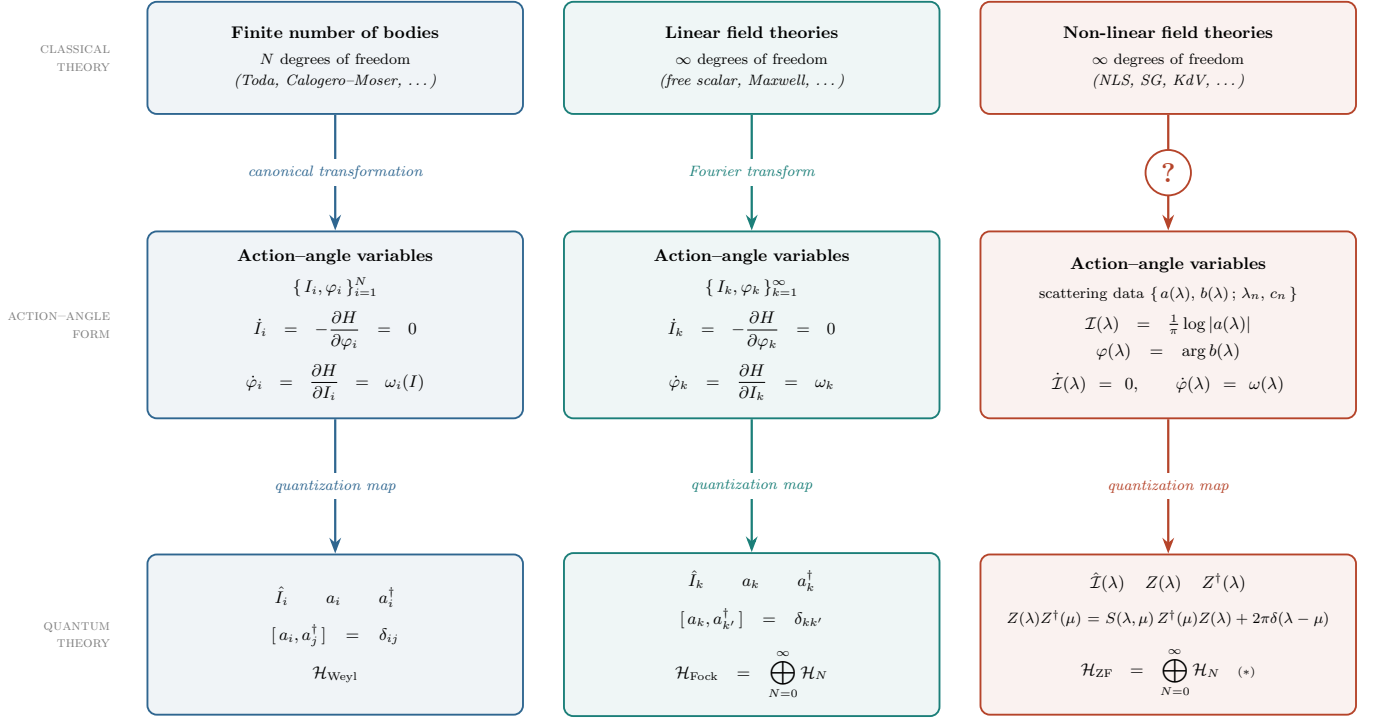

Quantization promotes the action--angle variables to operators, as is standard both for linear field theories and for finite-dimensional quantum integrable systems. The actions become the conserved charges, among them the Hamiltonian and the momentum, while each angle variable yields a pair of creation and annihilation operators of particle-like modes with definite energy and momentum. The program is summarized in Fig.~\ref{fig:quantisation-scheme}, whose three columns correspond to finitely many bodies, linear field theories and non-linear field theories. In all three columns the passage to the quantum theory consists of a canonical transformation to action--angle variables followed by a quantization map. The nature of the target Hilbert space differs between theories with finitely many bodies and field theories.

For finitely many bodies, the first arrow is an ordinary canonical transformation, and the Stone--Von Neumann theorem fixes the Hilbert space $\mathcal{H}_{\mathrm{Weyl}}$. For a linear field theory the first arrow is the Fourier transform, and the creation and annihilation operators generate the Fock space from a vacuum defined by a linear condition. For a non-linear field theory the question mark on the first arrow is answered by the ISM, which maps the non-linear problem to scattering data evolving according to linear ODEs, and
which reduces to the Fourier transform at linear order \cite{Ablowitz:1991xb}. The scattering data $\{a(\lambda),b(\lambda),\lambda_n,c_n\}$ are collected in the monodromy matrix: a continuum of actions $\mathcal{I}(\lambda)=\frac{1}{\pi}\log|a(\lambda)|$ labeled by the spectral parameter, a discrete set attached to the zeros $\lambda_n$ of $a(\lambda)$, and the conjugate angles given by the phases of $b(\lambda)$ and of the $c_n$ \cite{Faddeev:1987ph,Babelon:2003qtg}. The generating function of the local conserved charges is $\log a(\lambda)$, whose expansion at the singular points of the spectral plane produces the whole tower of charges. The monodromy matrix, and not the field, is therefore the natural variable of the theory: it solves the dynamics and it carries the Poisson structure through the classical $r$-matrix, which becomes the $R$-matrix of the commutation relations after quantization.

In the quantum theory the operators $Z(\lambda)$, $Z^{\dagger}(\lambda)$ play the role of ladder operators, but they are built non-linearly from the entries $A(\lambda)$, $B(\lambda)$ of the quantum monodromy matrix and satisfy the Zamolodchikov--Faddeev algebra, with the two-body $S$-matrix in place of the trivial commutation factor \cite{Wadati:1985cc,Smirnov:1992vz}. The existence of the vacuum, the normalization and completeness of the states, and the identification of the spectral parameter with a physical momentum are no longer self-evident and have to be established case by case.

\subsection{Non-relativistic and relativistic integrable theories}
\label{sec:number-operator}

The non-linear Schr\"odinger (NLS) equation and the SG equation sit in the same, rightmost column of Fig.~\ref{fig:quantisation-scheme} and are quantized by the same recipe, yet they differ in a single structural fact: the NLS model possesses a conserved number operator among its charges in involution, while the SG model, being relativistic, does not.

When the number operator is conserved, as in the NLS, the sectors of the Hilbert space with different particle number are dynamically separated, each with the Hamiltonian of a finite quantum system. For the NLS this is the Lieb--Liniger Hamiltonian of $N$ bosons with a contact interaction \cite{Lieb1963EXACTAO}, and the second-quantized theory merely packages the tower of fixed-$N$ Hamiltonians into a single operator. The vacuum is defined, as in perturbation theory, as the state annihilated by the field; the Bethe states built on it are the asymptotic states \cite{Thacker:1980ei,Wadati:1985cc}; no discretization is needed; and no parameter is renormalized. This is the meaning of the asterisk in Fig.~\ref{fig:quantisation-scheme}: the direct sum over $N$-particle sectors describes the asymptotic states in both models, but only for the NLS it is a decomposition of the exact dynamics.

For the SG model the situation is radically different. In the absence of a conserved number operator, off-shell creation and annihilation of particles take place, and the asymptotic particles -- breathers, solitons and antisolitons -- cannot coincide with the quasiparticles that quantize the action--angle variables, whose number is conserved. The quasiparticles are constructed on a lattice and do not appear as asymptotic states in the continuum limit; the asymptotic states are combinations of them \cite{Sklyanin:1979pfu}. The vacuum is built in two steps: a ``spurious'' vacuum annihilated by the angle-type operator, which is an eigenvector of the Hamiltonian but not the state of lowest energy, and the physical vacuum obtained from it by filling the negative-energy quasiparticles. Moreover, the renormalization of the parameters, absent in the NLS, emerges from the continuum limit of the lattice model as a consequence of the non-trivial structure of the vacuum. Finally, the angle operator carries zero topological charge: acting on a neutral vacuum, it creates only states in the zero-charge sector.

\subsection{The sine-Gordon model on the lattice}
\label{sec:lattice-sg}

We consider the SG equation for the quantum field $\phi(x,t)$,
\begin{align}
    &\phi_{tt}-\phi_{xx}+\tfrac{8m^2}{\beta}\sin(2\beta\phi)=0, \nonumber\\
    &[\phi(x),\phi_t(y)]=i\delta(x-y),
\end{align}
in the conventions of \cite{Babelon:2003qtg}, which differ from those of \cite{Sklyanin:1979pfu} by $m\rightarrow 4m$, $\beta\rightarrow 2\beta$. In terms of $\psi=2\beta\phi$,
\begin{align}
    &\psi_{tt}-\psi_{xx}+16m^2\sin\psi=0, \nonumber\\
    &[\psi(x),\psi_t(y)]=4i\beta^2\delta(x-y), \label{SGpsi}
\end{align}
so that $4m$ is the mass of the linearized theory, $\beta\rightarrow0$ is the semiclassical limit, and the vertex operators $e^{\pm i\psi}$ have scaling dimension $\Delta=\beta^2/\pi$ at the free massless point. Classically, the monodromy matrix on the line has the form $\left(\begin{smallmatrix} a & b\\ -\bar b & \bar a\end{smallmatrix}\right)$, and $\log a(\lambda)$ can be expanded at the two essential singularities of the spectral parameter,
\begin{align}
\log a(\lambda)&=\sum_{n=1}^{\infty}I^{+}_{n}\,(i\lambda)^{-n}, && \lambda\to\infty, \nonumber\\
\log a(\lambda)&=\sum_{n=1}^{\infty}I^{-}_{n}\,(i\lambda)^{n}, && \lambda\to 0, \label{expansions}
\end{align}
in the zero-charge sector, and the first coefficients give the Hamiltonian and the momentum \cite{Babelon:2003qtg},
\begin{align}
    H=\frac{2m}{\beta^2}\left(I_1^-+I_{1}^+\right), \quad
    P=\frac{2m}{\beta^2}\left(I_1^--I_{1}^+\right). \label{HPclassical}
\end{align}
The soliton solutions have mass $M_S=8m/\beta^2$ \cite{Babelon:2003qtg,Ablowitz:1973fn}.

Following \cite{Sklyanin:1979pfu}, the quantum model is regularized on a lattice with sites $x_n=-L+n\delta$, $n=0,\dots,N$, $N=2L/\delta$, in terms of the variables
\begin{align}
    \nu_n&=\delta \exp\left\{\frac{i}{2\delta}\int^{x_n+\delta}_{x_n}\psi\, dx\right\}, \nonumber\\
    \psi_n&= \exp\left\{-\frac{i}{4}\int^{x_n+\delta}_{x_n}\psi_t\, dx\right\},
\end{align}
which regularize the vertex operators and satisfy the Weyl relations
\begin{align}
\psi_n \nu_n = e^{-i\frac{\beta^2}{2}} \nu_n \psi_n,\qquad \psi_n \bar{\nu}_n = e^{i\frac{\beta^2}{2}} \bar{\nu}_n \psi_n, \label{commutation}
\end{align}
while operators on different sites commute. The transition matrix from site $n$ to site $n+1$ and the monodromy matrix are
\begin{align}
    U_n(\lambda)&=\begin{pmatrix}
        \psi_n &m\left(\lambda \bar{\nu}_n-\tfrac{1}{\lambda}\nu_n\right)\\m\left(\tfrac{1}{\lambda}\bar{\nu}_n-\lambda \nu_n\right)&\bar{\psi}_n
    \end{pmatrix}, \nonumber\\
    T_L(\lambda)&=\overset{\leftarrow}{\prod_n} U_n(\lambda)=\begin{pmatrix}
        A_L(\lambda) & B_L(\lambda) \\ -\bar{B}_L(\bar{\lambda})& \bar{A}_L(\bar{\lambda})
    \end{pmatrix}.
\end{align}
Since SG admits an ultralocal gauge, in which no derivatives of the canonical variables appear, the $U_n$ on different sites commute; hence a solution of $R(\lambda,\mu)\,U_n(\lambda)\otimes U_n(\mu)=U_n(\mu)\otimes U_n(\lambda)\,R(\lambda,\mu)$ implies the same relation for $T_L$, that is, the commutation relations of the action--angle operators. In the rapidities $\alpha$ and $\theta$, defined as $\lambda=e^{\alpha}$, $\mu=e^{\theta}$, the non-trivial entries of $R$ are \cite{Sklyanin:1979pfu}
\begin{align}
   c(\alpha,\theta)&=\frac{\sinh(\alpha-\theta)}{\sinh(\alpha-\theta+i\frac{\beta^2}{2})}, \nonumber\\
    b(\alpha,\theta)&=\frac{\sinh\left(i\frac{\beta^2}{2}\right)}{\sinh(\alpha-\theta+i\frac{\beta^2}{2})}, \label{c}
\end{align}
with $R_{11}=R_{44}=1$, $R_{22}=R_{33}=b$ and $R_{23}=R_{32}=c$.

In \cite{Sklyanin:1979pfu} it is shown that there exists a state $\ket{\Omega}$ such that 
\begin{align}
    &\bar{B}_L(\bar{\lambda})\ket{\Omega}=0,\qquad A_L(\lambda)\ket{\Omega}=e^{a(\alpha)L}\ket{\Omega}, \nonumber\\
    &a(\alpha)=2m^2\delta^2\cosh\left(2\alpha-i\frac{\beta^2}{2}\right). \label{a(lambda)}
\end{align}
This holds only at order $\delta^2$ \cite{Sklyanin:1979pfu}; however, if the lattice model is integrable at leading order in $\delta$, its continuum limit is the integrable
theory. A caveat will be included in Sec.~\ref{sec:range}. The states $\prod_{l=1}^{n}B_L(\lambda_l)\ket{\Omega}$ are eigenvectors of $\operatorname{Tr}T_L(\lambda)$ with eigenvalues
\begin{align}
    \Lambda(\lambda;\lambda_1,\dots,\lambda_n)={}&e^{a(\lambda)L}\prod_{l=1}^n \frac{1}{c(\lambda_l,\lambda)} \nonumber\\
    &+e^{\bar{a}(\bar{\lambda})L}\prod_{l=1}^n \frac{1}{c(\lambda,\lambda_l)} \label{eigenvaluesss}
\end{align}
provided the rapidities satisfy the Bethe equations \cite{Sklyanin:1979pfu}
\begin{align}
    e^{-im_{\delta} \sin \left(\frac{\beta^2}{2}\right) \sinh (2\alpha_l)L}&=\prod_{\underset{j\neq l}{j=1}}^n \frac{\sinh(\alpha_l-\alpha_j+i\frac{\beta^2}{2})}{\sinh(\alpha_l-\alpha_j-i\frac{\beta^2}{2})}, \nonumber\\
    m_\delta&=4m^2\delta. \label{Bethe}
\end{align}
They describe quasiparticles with momentum and energy
\begin{align}
    k'=m_\delta \sin \left(\frac{\beta^2}{2}\right) \sinh (2\alpha),\nonumber\\ E'=m_\delta \sin \left(\frac{\beta^2}{2}\right) \cosh (2\alpha), \label{kprime}
\end{align}
whose scattering is described by the two-body phase $S(\alpha)=\sinh(\alpha+i\frac{\beta^2}{2})/\sinh(\alpha-i\frac{\beta^2}{2})=e^{i\Phi(\alpha)}$. The spectral parameter $\alpha$ can be restricted to the strip $0\leq\operatorname{Im}\alpha\leq\pi/2$ \cite{Babelon:2003qtg,Sklyanin:1979pfu}, on whose boundaries the energies are real. For $0<\beta^2<2\pi$ the quasiparticles with $\operatorname{Im}\alpha=0$ have positive energy and those with $\operatorname{Im}\alpha=\pi/2$ negative energy. Since the negative energy
particles are unphysical, the lowest energy state of the
theory, namely the physical vacuum, is obtained by
filling the spurious vacuum with $\mathcal{M}$, a maximal set of non-coincident solutions of Eq.~\eqref{Bethe} with $\operatorname{Im}\alpha=\pi/2$
\begin{align}
    \ket{0}_{L,\delta}= \prod_{\alpha_l \in \mathcal{M}}B_L(\alpha_l)\ket{\Omega}.\label{unpolarized}
\end{align}
At $\beta^2=2\pi$ the factor $\sin(\beta^2/2)$ changes sign and the state to be filled changes discontinuously. This is the lattice counterpart of the bound of \cite{Coleman:1974bu}, beyond which the normal-ordered Hamiltonian is unbounded from below, and of the Berezinskii--Kosterlitz--Thouless transition \cite{Amit1980}. We therefore restrict the theory to $0<\beta^2<2\pi$. This restriction is discussed in more detail in Sec.~\ref{sec:range}.

\begin{figure}[htbp]
\centering
\resizebox{\columnwidth}{!}{%
\begin{tikzpicture}[font=\small]

\fill[colGrey!6] (-4.0,-2.2) rectangle (4.0,2.2);

\draw[axs] (-4.35,-2.75) -- (-4.35,3.15) node[above, lbl] {$\operatorname{Im}\alpha$};
\foreach \y/\lb in {2.2/{\tfrac{\pi}{2}}, 0/0, -2.2/{-\tfrac{\pi}{2}}}{
  \draw[tk] (-4.45,\y) -- (-4.25,\y);
  \node[lbl, anchor=east] at (-4.5,\y) {$\lb$};}

\draw[imagevacuum] (-4.0,-2.2) -- (4.0,-2.2);
\draw[vacuumline]  (-4.0, 2.2) -- (4.0, 2.2);
\draw[realline, -{Stealth[length=4pt]}] (-4.0,0) -- (4.22,0);
\node[lbl, anchor=west] at (4.28,0) {$\operatorname{Re}\alpha$};

\foreach \x in {-3.600,-3.381,-3.148,-2.902,-2.641,-2.363,-2.070,-1.759,
                -1.432,-1.090,-0.735,-0.370,0.000,0.370,0.735,1.090,1.432,
                 1.759,2.070,2.363,2.641,2.902,3.148,3.381,3.600}
  \node[vacuummode] at (\x,2.2) {};

\draw[guide] (-3.6,2.55) -- (-3.6,1.85);
\draw[guide] ( 3.6,2.55) -- ( 3.6,1.85);
\node[lbl, anchor=north] at (-3.6,1.8) {$-\Lambda$};
\node[lbl, anchor=north] at ( 3.6,1.8) {$\Lambda$};

\node[lblB, anchor=west] at (-4.0,2.72)
  {vacuum: $\alpha_l=\theta_l+i\tfrac{\pi}{2}$, \ $E<0$, \ all filled};
\node[lblA, anchor=west] at (-4.0,-0.52)
  {\ $E>0$: \ no quasiparticle};

\end{tikzpicture}}
\caption{%
Rapidity content of the physical vacuum $\ket{0}_{L,\delta}$: the negative-energy
line is completely filled (crosses) between the cut-off $\pm\Lambda$ of
Eq.~\eqref{cutoff}, with the density of Eq.~\eqref{rholambda}, and the
positive-energy line carries no quasiparticle.}
\label{fig:poles- vacuum }
\end{figure}
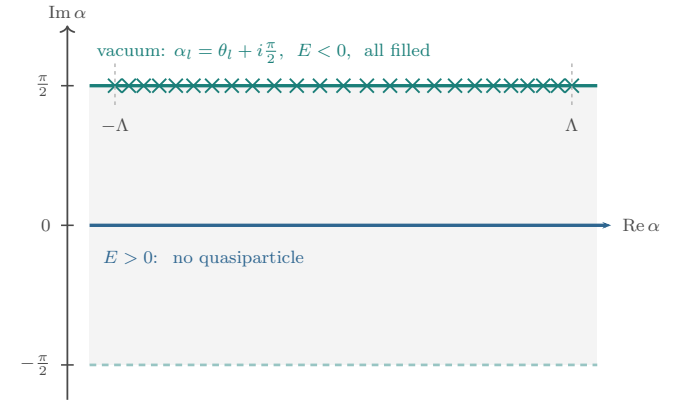

\subsection{Vacuum density and eigenvalues of the asymptotic states}
\label{sec:faddeev-results}

Writing $\alpha_l=\theta_l+i\frac{\pi}{2}$ and taking the logarithm of Eq.~\eqref{Bethe}, we obtain
\begin{align}
    m_\delta\sin \frac{\beta^2}{2} \sinh (2\theta_l)L=\sum_{j\neq l}\Phi(\theta_l-\theta_j)+2\pi l.\label{lnbethe}
\end{align}
The rapidities of the vacuum fill the interval $-\Lambda\leq\theta_l\leq\Lambda$, where the lattice cut-off $\Lambda$ is fixed by
\begin{align}
    m_\delta \sin \frac{\beta^2}{2}\sinh (2\Lambda)= \frac{2\pi}{\delta}. \label{cutoff}
\end{align}
As $L\rightarrow\infty$, the vacuum is described by the density $\frac{1}{(\alpha_{l+1}-\alpha_l)L}\rightarrow\rho_\Lambda(\alpha)$, which solves
\begin{align}
    2m_\delta\sin \frac{\beta^2}{2} \cosh (2\alpha)=2\pi\rho_\Lambda(\alpha)+\int_{-\Lambda}^\Lambda\Phi'(\alpha-\theta)\rho_\Lambda(\theta)d\theta, \label{lnbethesubtraction}
\end{align}
and for $\Lambda\rightarrow\infty$ the solution is \cite{Sklyanin:1979pfu}
\begin{align}
    \rho_\Lambda(\alpha)=Cm(4m\delta)^{p-1}\cosh p\alpha, \label{rholambda}
\end{align}
with
\begin{align}
    p&=\left\{ \begin{aligned}&\frac{2\pi}{2\pi-\beta^2}, && 0 < \beta^2 <\tfrac{4\pi}{3},\\ &\frac{4\pi}{\beta^2}, && \tfrac{4\pi}{3}< \beta^2 <2\pi,
    \end{aligned}\right.
    \\
    C&=16\left(\frac{16\pi}{\sin \frac{\beta^2}{2}}\right)^{-\frac{p}{2}} \nonumber\\ &\quad\times\left\{ \begin{aligned}&\frac{\cot \left( \frac{\pi\beta^2}{2(2\pi-\beta^2)} \right)}{\pi-\beta^2}, && 0 < \beta^2 <\tfrac{4\pi}{3},\\ &\frac{\tan \left( \frac{2\pi^2}{\beta^2} \right)}{2\pi-\beta^2}, && \tfrac{4\pi}{3}< \beta^2 <2\pi.
    \end{aligned}\right.\label{Cconstant}
\end{align}
The density depends on the lattice spacing $\delta$; this dependence is the subject of Sec.~\ref{sec:renormalization}. In the computations that follow, the steps that remove this dependence from the vacuum density are assumed to have been taken. The rapidity content of the vacuum is shown in Fig.~\ref{fig:poles- vacuum }.

The asymptotic states are built by adding rapidities on the line $\operatorname{Im}\alpha=0$ and letting the vacuum polarize around them \cite{Sklyanin:1979pfu}. Here polarization means that the added rapidities affects those of the vacuum, shifting and even removing some of them. Moreover, the eigenvalues associated with these
states are computed in terms of the shifted spectral parameter $\tau=\alpha-i\beta^2/4$, in which unitarity is manifest in the continuum limit. Finally, these eigenvalues
are obtained as the continuum limit of the eigenvalues of $\operatorname{Tr}T_L$ on the asymptotic states at finite $L,\delta$, divided by its eigenvalue on the vacuum $\ket{0}_{L,\delta}$, since only energy differences are physical. The construction of the zero-charge sector is summarized in Fig.~\ref{fig:hilbert-sg} below.

\begin{figure}[htbp]
\centering
\resizebox{\columnwidth}{!}{%
\begin{tikzpicture}[font=\small]

\fill[colGrey!6] (-4.0,-2.2) rectangle (4.0,2.2);

\draw[axs] (-4.35,-2.75) -- (-4.35,3.15) node[above, lbl] {$\operatorname{Im}\alpha$};
\foreach \y/\lb in {2.2/{\tfrac{\pi}{2}}, 0/0, -2.2/{-\tfrac{\pi}{2}}}{
  \draw[tk] (-4.45,\y) -- (-4.25,\y);
  \node[lbl, anchor=east] at (-4.5,\y) {$\lb$};}

\draw[imagevacuum] (-4.0,-2.2) -- (4.0,-2.2);
\draw[vacuumline]  (-4.0, 2.2) -- (4.0, 2.2);
\draw[realline, -{Stealth[length=4pt]}] (-4.0,0) -- (4.22,0);
\node[lbl, anchor=west] at (4.28,0) {$\operatorname{Re}\alpha$};

\foreach \x in {-3.600,-3.381,-3.148,-2.902,-2.641,-2.363,-2.070,-1.759,
                -1.432,-1.090,-0.665,-0.280,0.100,0.440,1.180,1.530,1.840,
                 2.130,2.400,2.660,2.915,3.155,3.385,3.600}
  \node[vacuummode] at (\x,2.2) {};
\node[vacuumhole] at (0.735,2.2) {};

\node[pole] (p1) at (0.735,0) {};
\draw[chain] (0.735,0.16) -- (0.735,2.02);

\draw[bkflow] (-1.55,1.62) -- (-0.15,1.62);
\draw[bkflow] ( 3.05,1.62) -- ( 1.65,1.62);
\node[lblB, anchor=east] at (-1.65,1.62) {$\theta'_l>\theta_l$};
\node[lblB, anchor=west] at ( 3.15,1.62) {$\theta'_l<\theta_l$};

\node[lblB, anchor=west] at (-4.0,2.72)
  {polarized vacuum: one mode removed near $\alpha_1$, the others shift};
\node[lblC, anchor=west] at (0.90,-0.42) {$\alpha_1\in I_1$, \ $\operatorname{Im}\alpha_1=0$};

\end{tikzpicture}}
\caption{%
Pole structure of the fundamental excitation $\ket{\alpha_1}_{L,\delta}$ of
Eq.~\eqref{1P}: a single rapidity on the positive-energy line, one hole in the vacuum
beside it, and the shift of the remaining vacuum rapidities.}
\label{fig:poles-first-breather}
\end{figure}
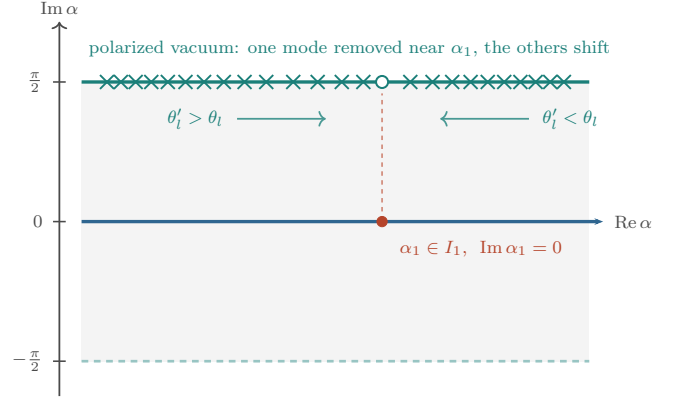

The fundamental excitation,
\begin{align}
    \ket{\alpha_1}_{L,\delta}=B_L(\alpha_1)\prod_{\theta'_l \in \mathcal{M}^*}B_L(\theta'_l+i\tfrac{\pi}{2})\ket{\Omega}, \label{1P}
\end{align}
contains one real rapidity $\alpha_1$; the set $\mathcal{M}^*$ of vacuum rapidities is polarized by the repulsion between the particle and the holes, with one hole opened next to $\alpha_1$ and the remaining rapidities displaced towards it (Fig.~\ref{fig:poles-first-breather}). Its eigenvalue is \cite{Sklyanin:1979pfu}
\begin{align}
    A_1(\tau;\alpha_1)=\frac{\sinh \left(\frac{2\pi(\tau-\alpha_1)}{2\pi-\beta^2}\right)-i\sin \left(\frac{\pi\beta^2}{2(2\pi-\beta^2)}\right)}{\sinh \left(\frac{2\pi(\tau-\alpha_1)}{2\pi-\beta^2}\right)+i\sin \left(\frac{\pi\beta^2}{2(2\pi-\beta^2)}\right)}, \label{A1result}
\end{align}
and the eigenvalue of a state with several fundamental excitations is the product of the single-particle factors.

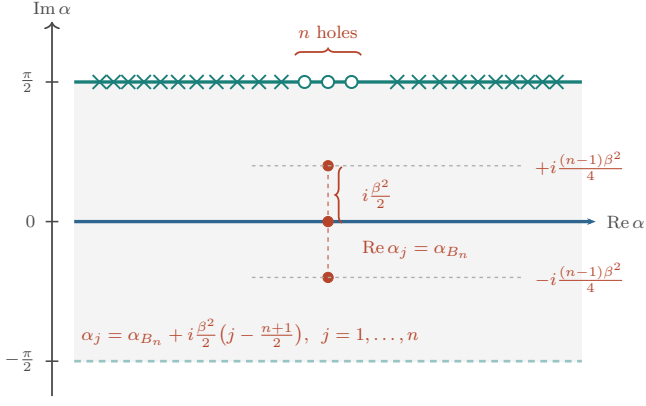
\begin{figure}[htbp]
\centering
\resizebox{\columnwidth}{!}{%
\begin{tikzpicture}[font=\small]

\fill[colGrey!6] (-4.0,-2.2) rectangle (4.0,2.2);

\draw[axs] (-4.35,-2.75) -- (-4.35,3.15) node[above, lbl] {$\operatorname{Im}\alpha$};
\foreach \y/\lb in {2.2/{\tfrac{\pi}{2}}, 0/0, -2.2/{-\tfrac{\pi}{2}}}{
  \draw[tk] (-4.45,\y) -- (-4.25,\y);
  \node[lbl, anchor=east] at (-4.5,\y) {$\lb$};}

\draw[imagevacuum] (-4.0,-2.2) -- (4.0,-2.2);
\draw[vacuumline]  (-4.0, 2.2) -- (4.0, 2.2);
\draw[realline, -{Stealth[length=4pt]}] (-4.0,0) -- (4.22,0);
\node[lbl, anchor=west] at (4.28,0) {$\operatorname{Re}\alpha$};

\foreach \x in {-3.600,-3.381,-3.148,-2.902,-2.641,-2.363,-2.070,-1.759,
                -1.432,-1.090,-0.735,1.090,1.432,1.759,2.070,2.363,2.641,
                 2.902,3.148,3.381,3.600}
  \node[vacuummode] at (\x,2.2) {};
\foreach \x in {-0.370,0.000,0.370} \node[vacuumhole] at (\x,2.2) {};
\draw[brc] (-0.52,2.62) -- (0.52,2.62);
\node[lblC, anchor=south] at (0,2.78) {$n$ holes};

\draw[chain] (0,-0.88) -- (0,0.88);
\foreach \y in {-0.88,0,0.88} \node[pole] at (0,\y) {};
\draw[brc] (0.22,-0.02) -- (0.22,0.86);
\node[lblC, anchor=west] at (0.42,0.44) {$i\tfrac{\beta^2}{2}$};

\draw[guide] (-1.2,0.88) -- (3.1,0.88);
\node[lblC, anchor=west] at (3.15,0.88) {$+i\tfrac{(n-1)\beta^2}{4}$};
\draw[guide] (-1.2,-0.88) -- (3.1,-0.88);
\node[lblC, anchor=west] at (3.15,-0.88) {$-i\tfrac{(n-1)\beta^2}{4}$};

\node[lblC, anchor=west] at (0.42,-0.44) {$\operatorname{Re}\alpha_j=\alpha_{B_n}$};
\node[lblC, anchor=west] at (-4.0,-1.75)
  {$\alpha_j=\alpha_{B_n}+i\tfrac{\beta^2}{2}\!\left(j-\tfrac{n+1}{2}\right)$, \ $j=1,\dots,n$};

\end{tikzpicture}}
\caption{%
Pole structure of the $n$-th breather $\ket{\alpha_{B_n}}_{L,\delta}$, drawn for
$n=3$ at $\beta^2=2\pi/5$: a chain of $n$ rapidities spaced by $i\beta^2/2$ and
symmetric about the real axis, together with the $n$ holes it opens in the vacuum.}
\label{fig:poles-breather}
\end{figure}

Bound states of $n$ quasiparticles correspond to strings of rapidities
\begin{align}
\alpha_j=\alpha_{B_n}+i\tfrac{\beta^2}{2}\left(j-\tfrac{n+1}{2}\right),\qquad j=1,\dots,n, \label{string}
\end{align}
spaced by $i\beta^2/2$, as required by the poles of $S(\alpha)$, symmetric about the real axis owing to the conjugation symmetry of the monodromy matrix, and with $n$ holes opened in the vacuum (Fig.~\ref{fig:poles-breather}). The number of such states is bounded, $1\leq n\leq n_{max}$ with $n_{max}<\frac{2\pi}{\beta^2}-1$, so that they exist only for $0<\beta^2<\pi$. Their eigenvalues are \cite{Sklyanin:1979pfu}
\begin{align}
    A_n(\tau;\alpha_{B_n})=\frac{\sinh \left(\frac{2\pi(\tau-\alpha_{B_n})}{2\pi-\beta^2}\right)-i\sin \left(\frac{n\pi\beta^2}{2(2\pi-\beta^2)}\right)}{\sinh \left(\frac{2\pi(\tau-\alpha_{B_n})}{2\pi-\beta^2}\right)+i\sin \left(\frac{n\pi\beta^2}{2(2\pi-\beta^2)}\right)}. \label{breathers}
\end{align}
As $\beta\rightarrow0$ the strings become the zeros of $a(\lambda)$ associated with the classical breathers, while the holes reproduce the classical solitons and antisolitons \cite{Babelon:2003qtg,Ablowitz:1973fn}.

\begin{figure}[!b]
\centering
\resizebox{\columnwidth}{!}{%
\begin{tikzpicture}[font=\small]

\fill[colGrey!6] (-4.0,-2.2) rectangle (4.0,2.2);

\draw[axs] (-4.35,-2.75) -- (-4.35,3.15) node[above, lbl] {$\operatorname{Im}\alpha$};
\foreach \y/\lb in {2.2/{\tfrac{\pi}{2}}, 0/0, -2.2/{-\tfrac{\pi}{2}}}{
  \draw[tk] (-4.45,\y) -- (-4.25,\y);
  \node[lbl, anchor=east] at (-4.5,\y) {$\lb$};}

\draw[imagevacuum] (-4.0,-2.2) -- (4.0,-2.2);
\draw[vacuumline]  (-4.0, 2.2) -- (4.0, 2.2);
\draw[realline, -{Stealth[length=4pt]}] (-4.0,0) -- (4.22,0);
\node[lbl, anchor=west] at (4.28,0) {$\operatorname{Re}\alpha$};

\foreach \x in {-3.600,-3.381,-3.148,-2.902,-2.641,-2.363,-1.759,
                -1.432,-1.090,-0.735,1.090,1.432,1.759,2.641,
                 2.902,3.148,3.381,3.600}
  \node[vacuummode] at (\x,2.2) {};
\foreach \x in {-0.370,0.000,0.370} \node[vacuumhole] at (\x,2.2) {};
\node[solhole] at (-2.070,2.2) {};
\node[solhole] at ( 2.070,2.2) {};

\draw[brc] (-0.52,2.62) -- (0.52,2.62);
\node[lblC, anchor=south] at (0,2.78) {$n-2$ holes};
\node[lblA, anchor=south] at (-2.070,2.48) {$\alpha_s$};
\node[lblA, anchor=south] at ( 2.070,2.48) {$\alpha_{\bar{s}}$};

\draw[chain] (0,-1.76) -- (0,1.76);
\foreach \y in {-1.76,-0.88,0,0.88,1.76} \node[pole] at (0,\y) {};
\draw[brc] (0.22,-0.02) -- (0.22,0.86);
\node[lblC, anchor=west] at (0.42,0.44) {$i\tfrac{\beta^2}{2}$};

\node[lblC, anchor=west] at (0.42,-0.44)
  {$\operatorname{Re}\alpha_j=\tfrac{\alpha_s+\alpha_{\bar{s}}}{2}$};

\end{tikzpicture}}
\caption{%
Pole structure of the soliton--antisoliton state
$\ket{\alpha_s,\alpha_{\bar{s}}}_{L,\delta}$, drawn for $n=n_{max}+2=5$ at
$\beta^2=2\pi/5$: the longest string the theory admits, with only $n-2$ holes beside
it and two independent holes at $\alpha_s$ and $\alpha_{\bar{s}}$.}
\label{fig:poles-solitons}
\end{figure}
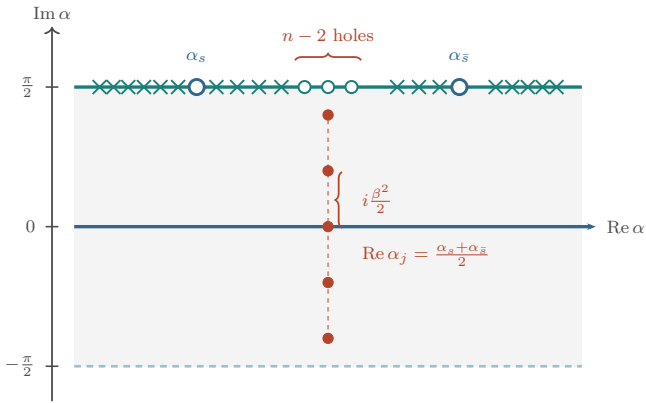

The soliton--antisoliton states contain charged particles. Their construction uses the duality with the massive Thirring model, in which adding particles to the longest string of neutral states produces charged states \cite{Korepin1980Thirring}. They correspond to a string longer than any breather string, $n=n_{max}+1$ or $n=n_{max}+2$ (the symmetric and antisymmetric combinations of two spin-$1/2$ particles), with a different polarization rule: $n-2$ holes next to the string and two holes at the rapidities $\alpha_s$ and $\alpha_{\bar s}$, which carry the two asymptotic particles (Fig.~\ref{fig:poles-solitons}). The eigenvalue for the symmetric and antisymmetric cases is \cite{Sklyanin:1979pfu}
\begin{align}
    A_{s\bar{s}}(\tau;\alpha_s,\alpha_{\bar{s}})=\prod_{\alpha=\alpha_s,\alpha_{\bar s}}\frac{\sinh\left(\frac{\pi(\tau-\alpha)}{2\pi-\beta^2}-i\frac{\pi}{4}\right)}{\sinh\left(\frac{\pi(\tau-\alpha)}{2\pi-\beta^2}+i\frac{\pi}{4}\right)}. \label{solitonscattering}
\end{align}
It factorizes into one factor per particle, so that the eigenvalue of a single soliton or antisoliton is also recovered:
\begin{align}
    A_s(\tau;\alpha)=A_{\bar{s}}(\tau;\alpha)=\frac{\sinh\left(\frac{\pi(\tau-\alpha)}{2\pi-\beta^2}-i\frac{\pi}{4}\right)}{\sinh\left(\frac{\pi(\tau-\alpha)}{2\pi-\beta^2}+i\frac{\pi}{4}\right)}. \label{solitoneig}
\end{align}
The continuation of Eq.~\eqref{solitonscattering} to $\alpha_s=\bar{\alpha}_{\bar{s}}=\alpha_{B_n}\pm i\left(\pi+\frac{n-1}{4}\beta^2\right)$ gives Eq.~\eqref{breathers} \cite{Sklyanin:1979pfu}, which identifies the bound states of fundamental excitations as breathers, i.e., bound states of a soliton and an antisoliton, in agreement with the pole structure of the exact $S$-matrix \cite{Zamolodchikov:1977py}. 

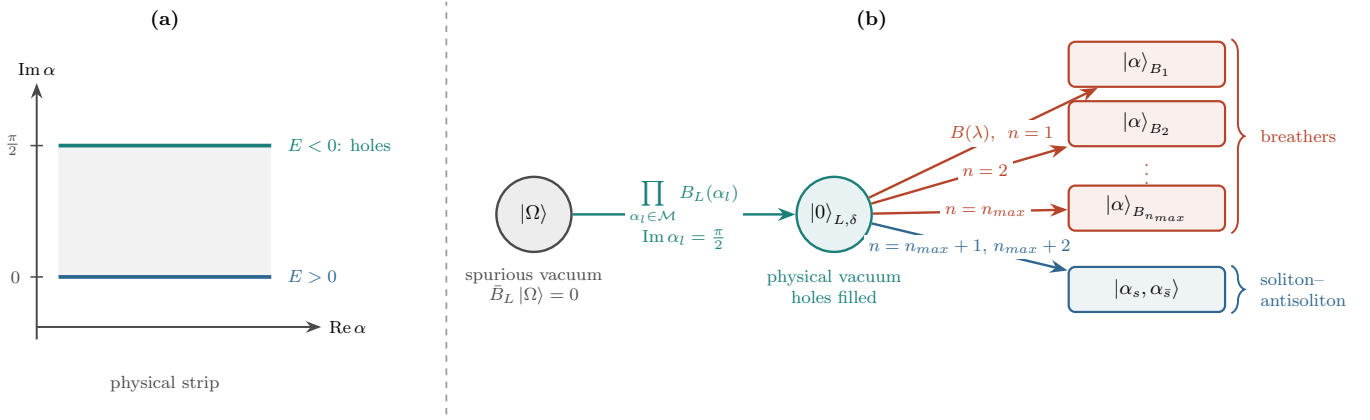
\begin{figure*}[t]
\centering
\resizebox{\textwidth}{!}{%
\begin{tikzpicture}[
  >=Stealth, font=\small,
  vac/.style={circle, draw=#1, line width=1.1pt, fill=#1!10,
              inner sep=0pt, minimum size=1.2cm},
  vac/.default=colGrey,
  st/.style={rounded corners=3pt, draw=#1, line width=1pt, fill=#1!8,
             inner sep=4pt, minimum width=2.5cm, minimum height=0.72cm},
  st/.default=colC,
  op/.style={->, line width=1.1pt, draw=#1},
  op/.default=colC,
  lb/.style={fill=white, inner sep=2pt, font=\footnotesize, text=#1},
  lb/.default=colC,
  hd/.style={font=\bfseries}
]

\node[hd] at (1.7,4.4) {(a)};
\fill[colGrey!7] (0,0.3) rectangle (3.4,2.4);
\draw[->, draw=colGrey, line width=0.9pt] (-0.35,-0.5) -- (4.2,-0.5) node[right,font=\footnotesize]{$\operatorname{Re}\alpha$};
\draw[->, draw=colGrey, line width=0.9pt] (-0.35,-0.7) -- (-0.35,3.4) node[above,font=\footnotesize]{$\operatorname{Im}\alpha$};
\draw[line width=1.7pt, draw=colA] (0,0.3) -- (3.4,0.3);
\draw[line width=1.7pt, draw=colB] (0,2.4) -- (3.4,2.4);
\draw[draw=colGrey, line width=0.7pt] (-0.45,0.3) -- (-0.25,0.3);
\draw[draw=colGrey, line width=0.7pt] (-0.45,2.4) -- (-0.25,2.4);
\node[font=\footnotesize, text=colGrey, anchor=east] at (-0.5,2.4) {$\tfrac{\pi}{2}$};
\node[font=\footnotesize, text=colGrey, anchor=east] at (-0.5,0.3) {$0$};
\node[font=\footnotesize, text=colA, anchor=west] at (3.55,0.3) {$E>0$};
\node[font=\footnotesize, text=colB, anchor=west] at (3.55,2.4) {$E<0$: holes};
\node[font=\footnotesize, align=center, text=colGrey] at (1.7,-1.4)
  {physical strip};

\draw[dashed, draw=colGrey!55, line width=0.8pt] (6.2,-1.9) -- (6.2,4.7);

\node[hd] at (13.0,4.4) {(b)};

\node[vac=colGrey] (om) at (7.6,1.3) {$\ket{\Omega}$};
\node[font=\footnotesize, align=center, text=colGrey] at (7.6,0.15)
  {spurious vacuum \\ $\bar{B}_L\ket{\Omega}=0$};

\node[vac=colB] (ph) at (12.4,1.3) {$\ket{0}_{L,\delta}$};
\node[font=\footnotesize, align=center, text=colB] at (12.4,0.15)
  {physical vacuum \\ holes filled};

\draw[op=colB] (om) -- node[lb=colB, align=center, pos=0.5]
  {$\displaystyle\prod_{\alpha_l\in\mathcal{M}}B_L(\alpha_l)$\\[1pt] $\operatorname{Im}\alpha_l=\tfrac{\pi}{2}$} (ph);

\node[st=colC] (b1) at (17.4,3.7) {$\ket{\alpha}_{B_1}$};
\node[st=colC] (b2) at (17.4,2.75) {$\ket{\alpha}_{B_2}$};
\node[st=colC] (bn) at (17.4,1.4) {$\ket{\alpha}_{B_{n_{max}}}$};
\node[font=\footnotesize, text=colC] at (17.4,2.05) {$\vdots$};
\node[st=colA] (ss) at (17.4,0.1) {$\ket{\alpha_s,\alpha_{\bar{s}}}$};

\draw[op=colC] (ph) -- node[lb=colC, pos=0.58] {$B(\lambda)$, \ $n=1$} (b1);
\draw[op=colC] (ph) -- node[lb=colC, pos=0.58] {$n=2$} (b2);
\draw[op=colC] (ph) -- node[lb=colC, pos=0.58] {$n=n_{max}$} (bn);
\draw[op=colA] (ph) -- node[lb=colA, pos=0.5] {$n=n_{max}+1,\,n_{max}+2$} (ss);

\draw[decorate, decoration={brace, amplitude=6pt}, draw=colC, line width=0.9pt]
  (18.75,4.1) -- (18.75,1.0);
\node[font=\footnotesize, text=colC, anchor=west] at (19.1,2.55) {breathers};
\draw[decorate, decoration={brace, amplitude=6pt}, draw=colA, line width=0.9pt]
  (18.75,0.48) -- (18.75,-0.28);
\node[font=\footnotesize, text=colA, align=left, anchor=west] at (19.1,0.1)
  {soliton--\\antisoliton};
\end{tikzpicture}}
\caption{%
\emph{(a)} Inside the strip $0\leq\operatorname{Im}\alpha\leq\pi/2$ the energies are real only on the two
boundaries: the quasiparticles on $\operatorname{Im}\alpha=0$ carry positive energy, those on
$\operatorname{Im}\alpha=\pi/2$ negative energy.
\emph{(b)} The spurious vacuum $\ket{\Omega}$, annihilated by $\bar{B}_L$, is therefore not the state of
lowest energy: filling its negative-energy holes with a maximal set $\mathcal{M}$ of solutions of
Eq.~\eqref{Bethe} gives the physical vacuum $\ket{0}_{L,\delta}$. Acting on $\ket{0}_{L,\delta}$ with the lattice creation operator $B_L$ on a string
of $n$ rapidities spaced by $i\beta^2/2$ and symmetric about the real axis gives the $n$-th breather,
up to $n_{max}<2\pi/\beta^2-1$; the two longest strings, with $n=n_{max}+1,n_{max}+2$, give, through a different polarization of the vacuum, the
symmetric and antisymmetric soliton--antisoliton states. For the fundamental excitation, the creation operator $B$ of the continuum model is constructed explicitly; the operators for the other states in the continuum model are not discussed. The eigenvalues of $\operatorname{Tr}T$ in the continuum model are the information relevant for what follows.}
\label{fig:hilbert-sg}
\end{figure*}

\section{Renormalization}
\label{sec:renormalization}

\subsection{The uniform continuum limit}
\label{sec:uniform-limit}

The density of Eq.~\eqref{rholambda} depends explicitly on $\delta$, and hence, through Eq.~\eqref{cutoff}, on the lattice cut-off $\Lambda$: the limit $\frac{1}{(\alpha_{l+1}-\alpha_l)L}\rightarrow\rho_\Lambda(\alpha)$ is not uniform in $\delta$. As observed in \cite{Sklyanin:1979pfu}, the appropriate variable is read off from $\rho_\Lambda(\alpha)$ itself,
\begin{align}
    L'=C\,L\,(4m\delta)^{p-1},\qquad \frac{1}{(\alpha_{l+1}-\alpha_l)L'} \underset{L'\rightarrow \infty}{\longrightarrow} \rho(\alpha),
\end{align}
and the continuum and infinite-volume limits give
\begin{align}
    \rho(\alpha)=m\cosh p\alpha. \label{vacuumdensity}
\end{align}
After the rescaling of Eq.~\eqref{vacuumdensity} below, in the relevant interval $0<\beta^2<\pi$, we obtain
\begin{align}
    \rho(\alpha)=m\cosh \alpha.
\end{align}
Since $L'$ depends on the product $m\delta$, the mass parameter $m$, which is the only renormalized parameter, is defined up to a finite multiplicative constant \cite{Sklyanin:1979pfu}, in agreement with perturbation theory \cite{Coleman:1974bu}. The rest of this section develops this observation into a renormalization scheme.

\subsection{Monodromy operator and conserved charges}
\label{sec:monodromy-operator}

On a generic asymptotic state of the zero-charge sector, $A_L(\lambda)$ acts, in the continuum limit, as the product of the single-particle eigenvalues of Sec.~\ref{sec:faddeev-results} times its eigenvalue $\Lambda_L(\alpha)$ on the physical vacuum, which diverges. The continuum operator is defined by dividing by $\Lambda_L(\alpha)$,
\begin{align}
    A(\lambda):=\lim\frac{A_L\left(\lambda e^{i\beta^2/4}\right)}{\Lambda_L\left(\alpha+i\beta^2/4\right)},\qquad A(\lambda)\ket{0}=\ket{0}, \label{Adef}
\end{align}
where the spectral shift makes $A$ unitary, as discussed above, and the conserved charges Hermitian, while $\ket{0}$ denotes the continuum limit of $\ket{0}_{L,\delta}$. The charges obtained from the expansion of $\ln A$ vanish on the physical vacuum, and the normalization by $\Lambda_L$ subtracts the vacuum energy for all $0<\beta^2<2\pi$ \cite{Sklyanin:1979pfu}. The operator $B(\lambda)$ that creates a fundamental excitation with rapidity $\ln\lambda$ creates a single-particle state for $0<\beta^2<\pi$; beyond $\beta^2=\pi$ the pole that gives the mass of the fundamental excitation gets a negative residue. The two operators satisfy $A(\lambda)B(\mu)=A_1(\lambda;\mu)\,B(\mu)A(\lambda)$, which reduces to the classical Poisson brackets of the scattering data as $\beta\rightarrow0$ \cite{Sklyanin:1979pfu,Babelon:2003qtg}.

All the eigenvalues of Sec.~\ref{sec:faddeev-results} depend on the rapidities through the combination $\frac{2\pi}{2\pi-\beta^2}(\tau-\alpha)$. We rescale the rapidities once more as
\begin{align}
    \alpha \rightarrow \frac{2\pi}{2\pi-\beta^2}\alpha. \label{rescaling}
\end{align}
This rescaling is a matter of convenience. The conserved charges are read off from the coefficients of $e^{\mp\theta}$ in the expansion of $\ln A$ at the essential singularities $\lambda=e^{\theta}\rightarrow \{0,\infty\}$, and classically $H$ and $P$ appear at first order in that expansion, Eqs.~\eqref{expansions} and~\eqref{HPclassical}. In the original variable the eigenvalues approach their asymptotic values as $e^{\mp\frac{2\pi}{2\pi-\beta^2}\theta}$, and their leading coefficients transform under a boost $\alpha\rightarrow\alpha+\chi$ as $e^{\pm\frac{2\pi}{2\pi-\beta^2}\chi}$; only in the rescaled variable do they transform as $e^{\pm\chi}$, like $H\mp P$. The rescaling therefore brings the quantum eigenvalues into the form of the classical expansion, with $H$ and $P$ at first order. From now on all formulas are understood with this rescaling applied.

With $\lambda=e^{\theta}$ and
\begin{align}
    \gamma=\frac{\pi\beta^2}{2\pi-\beta^2},
\end{align}
the logarithms of the eigenvalues on the one-particle states are
\begin{align}
    \ln A(\lambda)\ket{\alpha}_s&=\ln\left[\frac{\sinh\left(\frac{\theta-\alpha}{2}-i\frac{\pi}{4}\right)}{\sinh\left(\frac{\theta-\alpha}{2}+i\frac{\pi}{4}\right)}\right] \ket{\alpha}_s, \label{eigsolitons}\\
    \ln A(\lambda)\ket{\alpha}_{B_n}&=\ln\left[\frac{\sinh\left(\theta-\alpha\right)-i\sin\left(\frac{n\gamma}{2}\right)}{\sinh\left(\theta-\alpha\right)+i\sin\left(\frac{n\gamma}{2}\right)}\right] \ket{\alpha}_{B_n},\label{eigbreather}
\end{align}
with $n=1,\dots,n_{max}<\pi/\gamma$.

\subsection{The on-shell renormalization condition}
\label{sec:masses-renormalization}

The classical charges of Eq.~\eqref{HPclassical} carry the prefactor $2m/\beta^2$. The quantum construction does not fix this prefactor, for two reasons: the rescaling $L\rightarrow L'$ defines $m$ only up to a finite multiplicative constant \cite{Sklyanin:1979pfu}, and the ordering implicit in $\ln A$ is not specified. The two reasons are connected, as the renormalization of the mass is linked to the ordering ambiguity of the Hamiltonian \cite{Coleman:1974bu}. We therefore leave the prefactor free,
\begin{align}
    E=\frac{m}{2s(\beta)}\left(I_{1}^-+I_{1}^+\right), \quad
    P=\frac{m}{2s(\beta)}\left(I_{1}^--I_{1}^+\right), \label{Equantum}
\end{align}
with $s(\beta)\rightarrow\beta^2/4$ as $\beta\rightarrow0$, so that the classical expressions are recovered, and fix $s(\beta)$ below by a renormalization condition. Here $I_{1}^{\pm}$ are the first coefficients of the expansions of $\ln A_s$ or $\ln A_{B_n}$ at the two essential singularities,
\begin{align}
    I^-_{1}&=-i\lim_{\theta \rightarrow -\infty}e^{-\theta}\frac{d}{d\theta}\ln A, \nonumber\\
    I^+_{1}&=-i\lim_{\theta \rightarrow +\infty}e^{\theta}\frac{d}{d\theta}\ln A,
\end{align}
and the masses are $M=\frac{m}{s(\beta)}\sqrt{I^-_{1}I^+_{1}}$. From
\begin{align}
    \frac{d}{d\theta}\ln A_{s}&=\frac{i}{\cosh(\theta-\alpha)},\nonumber\\
    \frac{d}{d\theta}\ln A_{B_n}&=\frac{2i\cosh(\theta-\alpha)\sin\left(\frac{n\gamma}{2}\right)}{\sinh^2(\theta-\alpha)+\sin^2\left(\frac{n\gamma}{2}\right)}
\end{align}
we obtain $I^\mp_{1,s}=2e^{\mp\alpha}$ and $I^\mp_{1,B_n}=4\sin\left(\tfrac{n\gamma}{2}\right)e^{\mp\alpha}$, and hence
\begin{align}
    M_s=\frac{2m}{s(\beta)}, \qquad
    M_{B_n}=\frac{4m}{s(\beta)}\sin\left(\frac{n\gamma}{2}\right). \label{massesgeneric}
\end{align}

We impose the renormalization condition on the lightest particle of the spectrum. For $0<\beta^2<\pi$ this is the first breather, created by $B(\lambda)$, and we require its mass to equal the mass of the boson in the linearized version of Eq.~\eqref{SGpsi},
\begin{align}
    M_{B_1}=4m. \label{onshell}
\end{align}
This is an on-shell condition of pole-mass type: it fixes the position of a pole of the $S$-matrix in terms of the parameter of the equation of motion. No condition on the residue is needed, since neither the field nor $\beta$ is renormalized. The condition requires the first breather to belong to the spectrum, and it therefore defines the scheme in the interval $0<\beta^2<\pi$. Equation~\eqref{massesgeneric} gives $s(\beta)=\sin\left(\frac{\gamma}{2}\right)$, and therefore
\begin{align}
    M_s&=\frac{2m}{\sin\left(\frac{\gamma}{2}\right)}, \nonumber\\
    M_{B_n}&= \frac{4m}{\sin\left(\frac{\gamma}{2}\right)}\sin\left(\frac{n\gamma}{2}\right),\label{quantummasssoliton}
\end{align}
with $n=1,\dots,n_{max}$. Since $\gamma=\tfrac{1}{2}\beta^2+O(\beta^4)$, $\sin\left(\frac{\gamma}{2}\right)\rightarrow\frac{\beta^2}{4}$ as $\beta\rightarrow 0$, and $M_s$ reduces to the classical soliton mass $8m/\beta^2$. Under the substitution $\sin\left(\frac{\gamma}{2}\right)\rightarrow\frac{\gamma}{2}$ the masses coincide with the semiclassical results of \cite{Dashen:1975hd}, obtained independently from the path integral; the sine resums the perturbative series, of which $\frac{\gamma}{2}$ is the leading term. The ratios $M_{B_n}/M_s=2\sin\left(\frac{n\gamma}{2}\right)$ do not depend on the renormalization condition and agree with the pole structure of the exact $S$-matrix \cite{Zamolodchikov:1977py}. Once Eq.~\eqref{onshell} is imposed, the multiplicative ambiguity of $m$ is removed: $m$ is one quarter of the mass of a particle of the spectrum. At the upper end of the interval, $\beta^2=\pi$, one has $\gamma=\pi$ and $M_s=2m$, so that $M_{B_1}=2M_s$ and the first breather reaches the soliton--antisoliton threshold.

\subsection{The renormalization group and the dependence on the cut-off}
\label{sec:cutoff}

A comparison of Eq.~\eqref{rholambda} with Eq.~\eqref{vacuumdensity} shows that the entire dependence on the lattice spacing is carried by the ratio between the regularized box length and the length in which the continuum limit is uniform,
\begin{align}
    Z_L\equiv\frac{L'}{L}=\frac{\rho_\Lambda(\alpha)}{\rho(\alpha)}=C\,(4m\delta)^{p-1}. \label{ZL}
\end{align}
The factors $C(\beta^2)$ and $(4m\delta)^{p-1}$ play different roles. The exponent $p-1$ is universal and is the only information on the cut-off that the scheme retains. The constant $C(\beta^2)$ of Eq.~\eqref{Cconstant} is not universal and therefore depends on our truncation at order $\delta^2$; it is absorbed into $L'$.

The field is not rescaled, since the Weyl relations of Eq.~\eqref{commutation} are fixed by $\beta$, and $\beta$ is not renormalized \cite{Coleman:1974bu}. The mass parameter receives the finite factor $s(\beta)/(\beta^2/4)$, which does not depend on $\delta$. The renormalization-group (RG) differential equations for $(m,\beta)$ with respect to the scale $\delta$ are therefore trivial, with vanishing right-hand side.
 
The non-trivial part of the RG is contained in the following differential equation, which depends only on the universal part of $Z_L$
\begin{align}
    \delta\frac{\partial\ln Z_L}{\partial\delta}=p-1.
\end{align}
In the interval of the scheme, $0<\beta^2<\pi$, one has $p=2/(2-\Delta)$ with $\Delta=\beta^2/\pi$, so that $p-1=\Delta/(2-\Delta)$: the exponent is fixed by the scaling dimension $\Delta$ of the vertex operator. In a conventional scheme the same exponent controls the cut-off dependence on of the bare coefficient of the cosine at fixed physical mass. The information carried by the renormalization flow has therefore not disappeared: it has been moved from the coefficient $m$ to the box length $L$ before the continuum limit is taken. As $L'\rightarrow\infty$ this procedure yields a quantum field theory with a trivial flow: an RG fixed point for $(m,\beta)$. The second branch of Eq.~\eqref{Cconstant}, $\tfrac{4\pi}{3}<\beta^2<2\pi$, where $p=4\pi/\beta^2$ differs both from $2/(2-\Delta)$ and from the rescaling factor of Eq.~\eqref{rescaling}, lies outside the interval of the scheme.

\subsection{Comparison with Coleman's normal ordering}

In the scheme of \cite{Coleman:1974bu} the cosine is normal ordered with respect to a free field of mass $\mu$. Changing the ordering mass gives, up to $c$-numbers,
\begin{align}
    :\cos\psi:_{\mu}=\left(\frac{\mu'}{\mu}\right)^{\Delta}:\cos\psi:_{\mu'},
\end{align}
so the coefficient $\alpha_\mu$ (the equivalent of $m$) of $:\cos\psi:_\mu$ in the Hamiltonian obeys $\alpha_{\mu'}=\alpha_\mu(\mu'/\mu)^{\Delta}$. In this scheme the differential equation for $m$ is therefore not trivial: the parameter ''runs''. The combination $\alpha_\mu\mu^{-\Delta}$ is invariant under a rescaling of the subtraction scale $\mu$, and every mass has the form
\begin{align}
    M=\kappa(\beta)\left(\alpha_\mu\,\mu^{-\Delta}\right)^{\frac{1}{2-\Delta}}. \label{colemanmass}
\end{align}
No observable runs in any scheme: the explicit dependence on $\mu$ cancels against the running of $\alpha_\mu$. There are three differences with respect to the scheme presented above. (i) An observable is a function of a running parameter and of the subtraction scale, rather than of parameters that do not run; a self-consistent choice of $\mu$ in terms of $\alpha_\mu$, such as the one of \cite{Coleman:1974bu} that removes the tadpoles, eliminates $\mu$ and plays the role of our condition~\eqref{onshell}. (ii) Two different pieces of information enter Eq.~\eqref{colemanmass} and must be kept apart. The first is the relation between the physical scale and the parameter of the action, encoded in the overall normalization of $\kappa(\beta)$, which is supplied externally, either as a semiclassical series \cite{Dashen:1975hd} or exactly from integrability \cite{ZamolodchikovAl1995}. In our scheme its counterpart is the non-universal constant $C(\beta^2)$, which is absorbed into $Z_L$ and never needed. The second is the set of mass ratios, which normal ordering does not produce either: they have to be taken from the semiclassical spectrum \cite{Dashen:1975hd} or from the exact $S$-matrix \cite{Zamolodchikov:1977py}. Here they follow from the eigenvalues of the monodromy operator, Eq.~\eqref{massesgeneric}, within the same regularized construction that defines the theory, and only the overall scale is fixed by Eq.~\eqref{onshell}. This second point, and not the bypassing of the scale--parameter relation, is what distinguishes our scheme. (iii) For $\Delta\geq1$, that is, $\beta^2\geq\pi$, normal ordering no longer makes the vacuum energy density finite, and an additional subtraction is required \cite{Coleman:1974bu}. In the QISM construction this subtraction is performed by the normalization of $A(\lambda)$ in Eq.~\eqref{Adef}, for all $0<\beta^2<2\pi$; this is a property of the construction of the spectrum, which extends beyond the interval of the scheme.

\subsection{Comparison with conformal perturbation theory}

In conformal perturbation theory the model is defined as the free compact boson perturbed by the vertex operators, normalized conformally,
\begin{align}
    \langle e^{i\psi}(x)\,e^{-i\psi}(0)\rangle\underset{|x|\to0}{\longrightarrow}|x|^{-2\Delta},
\end{align}
with a coupling $\lambda$ of mass dimension $2-\Delta$ \cite{Zamolodchikov:1989ASPM,LukyanovZamolodchikov1997}. The dependence on the cut-off is absorbed into the normalization of the composite operator, not into the field. For $\Delta<1$ the expansion is ultraviolet finite and $\lambda$ does not run, and the masses are exact functions of $\lambda$,
\begin{align}
    M_s=\tilde{\kappa}(\beta)\,\lambda^{\frac{1}{2-\Delta}}, \label{cptmass}
\end{align}
with $\tilde{\kappa}(\beta)$ known in closed form \cite{ZamolodchikovAl1995,LukyanovZamolodchikov1997}. As far as the renormalization of the parameters is concerned, this parametrization is on the same footing as ours: in both, the parameters do not run and no subtraction scale appears. Two differences remain, however, and they are the same as in the previous case. (i) The function $\tilde{\kappa}(\beta)$, which relates the soliton mass to the dimensionful coupling $\lambda$, is not produced by conformal perturbation theory but obtained from exact results of integrability \cite{ZamolodchikovAl1995}. Nor are the mass ratios
produced by conformal perturbation theory at all; they are taken from the exact $S$-matrix \cite{Zamolodchikov:1977py}, whereas here they are eigenvalues of the monodromy operator computed within the same regularized construction. (ii) For $\Delta\geq1$ the vacuum energy of conformal perturbation theory is ultraviolet divergent and needs a subtraction, which the QISM construction performs automatically. The three parametrizations are summarized in Table~\ref{tab:schemes}.

\begin{table*}[t]
\caption{Three parametrizations of the quantum SG model, $\Delta=\beta^2/\pi$. All the rows refer to $0<\Delta<1$, the interval of the on-shell scheme, and highlight its key differences from
previously known methods.}
\label{tab:schemes}
\begin{ruledtabular}
\begin{tabular}{lccc}
 & \begin{tabular}[t]{@{}c@{}}Normal\\ordering \cite{Coleman:1974bu}\end{tabular} & \begin{tabular}[t]{@{}c@{}}Conformal perturbation\\theory \cite{Zamolodchikov:1989ASPM,LukyanovZamolodchikov1997}\end{tabular} & \begin{tabular}[t]{@{}c@{}}QISM,\\this work\end{tabular} \\
\hline
\begin{tabular}[t]{@{}l@{}}Cut-off dependence\end{tabular} & coefficient $\alpha_\mu$ of $:\cos\psi:_\mu$ & normalization of $e^{\pm i\psi}$ & box length, $Z_L$ \\
Parameters & $(\alpha_\mu,\mu,\beta)$, $\alpha_\mu$ runs & $(\lambda,\beta)$, no running & $(m,\beta)$, no running \\
Masses & $\kappa(\beta)\,(\alpha_\mu\mu^{-\Delta})^{1/(2-\Delta)}$ & $\tilde{\kappa}(\beta)\,\lambda^{1/(2-\Delta)}$ & Eq.~\eqref{quantummasssoliton} \\
Mass ratios & external \cite{Dashen:1975hd,Zamolodchikov:1977py} & external \cite{Dashen:1975hd,Zamolodchikov:1977py} & internal, Eq.~\eqref{massesgeneric} \\
Scale--parameter relation & \begin{tabular}[t]{@{}c@{}}$\kappa(\beta)$, \\external \cite{Dashen:1975hd,ZamolodchikovAl1995}\end{tabular} & \begin{tabular}[t]{@{}c@{}}$\tilde\kappa(\beta)$, \\external \cite{Dashen:1975hd,ZamolodchikovAl1995}\end{tabular} & \begin{tabular}[t]{@{}c@{}}$C(\beta^2)$ in $Z_L$,\\not needed, Eq.~\eqref{onshell}\end{tabular} \\
\end{tabular}
\end{ruledtabular}
\end{table*}

\subsection{Relation to other lattice regularizations}

Other exact regularizations of the model exist. The massive Thirring model (dual to the SG model \cite{Coleman:1974bu}) was solved by the coordinate Bethe ansatz with a cut-off on the bare rapidities, removed together with a renormalization of the bare mass \cite{Bergknoff1979PRD}. Lattice versions of the SG model that preserve the $R$-matrix structure at finite lattice spacing were constructed in \cite{Izergin1982}. In the light-cone lattice approach \cite{Destri1987} the continuum limit is reached by sending an inhomogeneity parameter to infinity together with the inverse lattice spacing, at fixed physical mass. In all these approaches the lattice is a device for defining the continuum theory, and renormalization, when discussed, takes the conventional form of a cut-off-dependent relation between a bare lattice parameter and the physical mass. The uniform limit of Sec.~\ref{sec:uniform-limit} places the dependence on the cut-off in $Z_L$, so that the parameters appearing in the continuum charges do not depend on it; supplemented by Eq.~\eqref{onshell}, this makes it a renormalization scheme rather than merely a regularization.

\subsection{Range of validity}
\label{sec:range}

Three intervals must be distinguished. The theory, with the eigenvalues of Sec.~\ref{sec:faddeev-results} and the subtraction of the vacuum energy, is valid in $0<\beta^2<2\pi$. The uniform limit with an exponent of $Z_L$ fixed by $\Delta$, and the rescaling of Eq.~\eqref{rescaling}, are derived here for $0<\beta^2<\tfrac{4\pi}{3}$, the first branch of Eq.~\eqref{Cconstant}. The renormalization scheme, defined by the on-shell condition~\eqref{onshell}, requires the first breather to belong to the spectrum and is therefore defined for $0<\beta^2<\pi$, inside that branch; within this interval all the results of this section hold.

For $\pi\leq\beta^2<2\pi$ the first breather is absent and the soliton is the lightest particle, so a different normalization condition would be needed. However, Eq.~\eqref{quantummasssoliton} cannot be continued to this interval: $\sin\left(\frac{\gamma}{2}\right)$ vanishes at $\gamma=2\pi$, that is, at $\beta^2=\tfrac{4\pi}{3}$, where $M_s$ at fixed $m$ diverges. In order to investigate the interval $\pi\leq\beta^2<2\pi$ properly, one should use the integrable lattice versions of \cite{Izergin1982}. In fact, not only does $C(\beta^2)$ depend on the order in $\delta$, but the point $\beta^2=4\pi/3$ will also shift towards $\beta^2=2\pi$ when higher orders in $\delta$ are considered. In lattice theory that is integrable at finite $\delta$, this point lies at $\beta^2=2\pi$ \cite{Izergin1982}. Hence the bound, $\beta^2=\tfrac{4\pi}{3}$, has no physical meaning: the soliton mass can be defined for $0\leq\beta^2<2\pi$ without diverging, and also the uniform limit can be defined in the same interval with the exponent of $Z_L$ still fixed by the same function of $\Delta$.

At $\beta^2=2\pi$ the vertex operator becomes marginal, $\Delta=2$, and the continuum theory reaches the BKT point \cite{Amit1980}. There $\gamma\rightarrow\infty$ and the constant $C$ of Eq.~\eqref{Cconstant} becomes indeterminate. Our analysis extracts only the universal part of the dependence on the cut-off in the interval of the scheme, and it does not describe the approach to the BKT point, which would require the full lattice model \cite{Izergin1982} beyond the leading form of the vacuum density; we do not address it here. Beyond this point the cosine is irrelevant and the continuum SG model is not a massive theory. The lattice Bethe equations remain well defined, but under $\beta^2\rightarrow\beta^2+2\pi$ one has $\sin\left(\frac{\beta^2}{2}\right)\rightarrow-\sin\left(\frac{\beta^2}{2}\right)$ and $c(\lambda,\mu)\rightarrow-c(\lambda,\mu)$, while $S(\alpha)$ is unchanged; combined with the shift $\alpha\rightarrow\alpha+i\frac{\pi}{2}$ of all rapidities, this maps Eqs.~\eqref{Bethe}, \eqref{a(lambda)} and~\eqref{eigenvaluesss} into themselves, up to an overall sign of the eigenvalue for an odd number of roots. The spectrum that the lattice produces for $2\pi<\beta^2<4\pi$ is therefore that of the model at $\beta^2-2\pi$, and it cannot interpreted as the SG theory beyond the BKT point.

\section{Conclusions}
\label{sec:conclusions}

The continuum and infinite-volume limits of the lattice SG model are not independent: they become uniform only after the length of the box is rescaled by the factor $Z_L$ of Eq.~\eqref{ZL} \cite{Sklyanin:1979pfu}. We have shown that this factor carries the entire dependence on the cut-off, with an exponent fixed by the scaling dimension of the vertex operator, and that, supplemented by the on-shell condition $M_{B_1}=4m$, it defines a renormalization scheme. Neither the field nor $\beta$ are renormalized, and $m$ receives a finite renormalization. The parameters do not run, no subtraction scale is introduced, and the masses of Eq.~\eqref{quantummasssoliton} follow from the eigenvalues of the monodromy operator as functions of $(m,\beta)$; they reduce to the classical soliton mass as $\beta\rightarrow0$ and agree with the semiclassical results of \cite{Dashen:1975hd} at first order in $\beta^2$. Two intervals are involved: (i) the theory theory, described
by its Hilbert space and by the eigenvalues of the monodromy matrix, extends to $0<\beta^2<2\pi$ if supplemented with a lattice theory valid at finite $\delta$ \cite{Izergin1982}; (ii) the scheme is defined for $0<\beta^2<\pi$, where the first breather belongs to the spectrum; at $\beta^2=\pi$ the soliton mass is $M_s=2m$ and the first breather reaches the soliton-antisoliton threshold.

With respect to Coleman's normal ordering \cite{Coleman:1974bu} and with conformal perturbation theory \cite{Zamolodchikov:1989ASPM,LukyanovZamolodchikov1997}, the difference lies in the parametrization, not in the physics: in no scheme does an observable depend on a renormalization scale. Two features are specific to our scheme. (i) The mass ratios, which the other two schemes do not compute but import from the exact $S$-matrix, are an output of the construction: they are eigenvalues of the monodromy operator on the asymptotic states, computed in the same regularization that defines the theory. (ii) The relation between the physical scale and the parameters, which in the other schemes is encoded in $\kappa(\beta)$ or $\tilde\kappa(\beta)$, corresponds here to the non-universal constant $C(\beta^2)$; it is absorbed into $Z_L$ together with the dependence on the cut-off and never needs to be computed. 

The comparison with the NLS model, where the action--angle quantization is structurally identical and no parameter is renormalized, shows that renormalization is a property of the regularization limit. This provides a natural motivation for
using the quantities involved in this limit to renormalize the theory.

The construction relies on one structural property of the SG model \cite{Sklyanin:1979pfu}, ultralocality, and we expect it to extend to all ultralocal integrable theories with a massive spectrum. Ultralocality is what allows the lattice regularization: the local transition operators on different sites commute, so that the commutation relations follow; the algebraic Bethe ansatz then yields the Bethe equations, from which the physical vacuum and all the Bethe states, the necessary ingredients of the construction, are computed explicitly.

An interesting extension of this method would be to asymptotically free theories, such as the non-linear sigma models \cite{Polyakov1975}, in which the
mass scale is generated by dimensional transmutation
and depends on the cut-off through a factor exponential in the inverse coupling. The dimensionless coupling is then not an independent parameter of the continuum theory, and, since the mass is the only scale, the entire dependence on the cut-off could still be absorbed into a factor multiplying a length: $Z_L$ would no longer be a power of $\delta$, and the only parameter left would be a mass fixed on shell. Remarkably, one would then obtain mass spectra as functions of parameters that do not run.

For the sigma models, the obstruction is precisely
ultralocality: their classical Poisson structure is non-
ultralocal \cite{Maillet1986}, and the construction cannot be set up directly \cite{Smirnov:1992vz}. Ultralocal lattice regularizations have, however, been constructed for non-ultralocal sigma models, in particular for the principal chiral field \cite{FaddeevReshetikhin1986}. Whether the vacuum density of these models has a uniform limit through a factor $Z_L$, and what form this factor takes, are questions left for future work.

We conclude by stressing two conceptual aspects. (i) We have given renormalization a microscopic interpretation: in theories with a single mass scale, the cut-off dependence
can be placed in the box length of the lattice model. (ii) A quantum field theory is commonly specified by its RG fixed points and by the flow between them \cite{WilsonKogut1974}; in this picture, the SG theory is a conformal field theory perturbed away from an ultraviolet fixed point \cite{Zamolodchikov:1989ASPM}. In our scheme, instead, the infrared data (the masses and the $S$-matrix) are eigenvalues of the monodromy operator, obtained without integrating a flow, and the ultraviolet behavior of the theory is regularized separately.

\begin{acknowledgments}
We acknowledge support from the HORIZON-ERC-2023-ADG HYPERSPIM project,
Grant No.~101139828.
\end{acknowledgments}

\bibliographystyle{apsrev4-2}
\bibliography{bibliografia}

@article{Ablowitz:1973fn,
    author  = "Ablowitz, M. J. and Kaup, D. J. and Newell, A. C. and Segur, H.",
    title   = "{Method for Solving the Sine--Gordon Equation}",
    doi     = "10.1103/PhysRevLett.30.1262",
    journal = "Phys. Rev. Lett.",
    volume  = "30",
    pages   = "1262--1264",
    year    = "1973"
}

@book{Ablowitz:1991xb,
    author    = "Ablowitz, M. J. and Clarkson, P. A.",
    title     = "{Solitons, Nonlinear Evolution Equations and Inverse Scattering}",
    publisher = "Cambridge University Press",
    address   = "Cambridge",
    series    = "London Mathematical Society Lecture Note Series",
    volume    = "149",
    doi       = "10.1017/CBO9780511623998",
    year      = "1991"
}

@article{Amit1980,
    author  = "Amit, Daniel J. and Goldschmidt, Yadin Y. and Grinstein, G.",
    title   = "{Renormalisation group analysis of the phase transition in the 2D Coulomb gas, Sine-Gordon theory and XY-model}",
    journal = "J. Phys. A",
    volume  = "13",
    number  = "2",
    pages   = "585--620",
    year    = "1980",
    doi     = "10.1088/0305-4470/13/2/024"
}

@book{Babelon:2003qtg,
    author    = "Babelon, Olivier and Bernard, Denis and Talon, Michel",
    title     = "{Introduction to Classical Integrable Systems}",
    doi       = "10.1017/CBO9780511535024",
    isbn      = "978-0-521-03670-2, 978-0-511-53502-4",
    publisher = "Cambridge University Press",
    address   = "Cambridge",
    series    = "Cambridge Monographs on Mathematical Physics",
    year      = "2003"
}

@article{Bergknoff1979PRD,
    author    = "Bergknoff, H. and Thacker, H. B.",
    title     = "{Structure and solution of the massive Thirring model}",
    journal   = "Phys. Rev. D",
    volume    = "19",
    number    = "12",
    pages     = "3666--3681",
    year      = "1979",
    publisher = "American Physical Society",
    doi       = "10.1103/PhysRevD.19.3666"
}

@article{Coleman:1974bu,
    author       = "Coleman, Sidney R.",
    editor       = "Stone, M.",
    title        = "{The Quantum Sine-Gordon Equation as the Massive Thirring Model}",
    reportNumber = "Print-74-1722 (HARVARD)",
    doi          = "10.1103/PhysRevD.11.2088",
    journal      = "Phys. Rev. D",
    volume       = "11",
    pages        = "2088",
    year         = "1975"
}

@article{Dashen:1975hd,
    author       = "Dashen, Roger F. and Hasslacher, Brosl and Neveu, Andre",
    title        = "{The Particle Spectrum in Model Field Theories from Semiclassical Functional Integral Techniques}",
    reportNumber = "COO-2220-37",
    doi          = "10.1103/PhysRevD.11.3424",
    journal      = "Phys. Rev. D",
    volume       = "11",
    pages        = "3424",
    year         = "1975"
}

@article{Destri1987,
    author  = "Destri, C. and de Vega, H. J.",
    title   = "{Light-cone lattice approach to fermionic theories in 2D: The massive Thirring model}",
    journal = "Nucl. Phys. B",
    volume  = "290",
    pages   = "363--391",
    year    = "1987",
    doi     = "10.1016/0550-3213(87)90193-3"
}

@book{Faddeev:1987ph,
    author    = "Faddeev, L. D. and Takhtajan, L. A.",
    title     = "{Hamiltonian Methods in the Theory of Solitons}",
    publisher = "Springer-Verlag",
    address   = "Berlin, Heidelberg",
    series    = "Springer Series in Soviet Mathematics",
    isbn      = "978-3-540-15579-9",
    doi       = "10.1007/978-3-540-69969-9",
    year      = "1987"
}

@article{FaddeevReshetikhin1986,
    author  = "Faddeev, L. D. and Reshetikhin, N. Yu.",
    title   = "{Integrability of the principal chiral field model in 1+1 dimension}",
    journal = "Ann. Phys.",
    volume  = "167",
    number  = "2",
    pages   = "227--256",
    year    = "1986",
    doi     = "10.1016/0003-4916(86)90201-0"
}

@article{Izergin1982,
    author  = "Izergin, A. G. and Korepin, V. E.",
    title   = "{Lattice versions of quantum field theory models in two dimensions}",
    journal = "Nucl. Phys. B",
    volume  = "205",
    pages   = "401--413",
    year    = "1982",
    doi     = "10.1016/0550-3213(82)90365-0"
}

@article{Korepin1980Thirring,
    author  = "Korepin, V. E.",
    title   = "{The Mass Spectrum and the S-Matrix of the Massive Thirring Model in the Repulsive Case}",
    journal = "Commun. Math. Phys.",
    volume  = "76",
    number  = "2",
    pages   = "165--176",
    year    = "1980",
    doi     = "10.1007/BF01212824"
}

@book{Korepin:1993kvr,
    author    = "Korepin, V. E. and Bogoliubov, N. M. and Izergin, A. G.",
    title     = "{Quantum Inverse Scattering Method and Correlation Functions}",
    doi       = "10.1017/CBO9780511628832",
    isbn      = "978-0-511-62883-2",
    publisher = "Cambridge University Press",
    address   = "Cambridge",
    series    = "Cambridge Monographs on Mathematical Physics",
    year      = "1993"
}

@article{Lieb1963EXACTAO,
    author  = "Lieb, Elliott H. and Liniger, Werner",
    title   = "{Exact Analysis of an Interacting Bose Gas. I. The General Solution and the Ground State}",
    journal = "Phys. Rev.",
    volume  = "130",
    pages   = "1605--1616",
    year    = "1963",
    doi     = "10.1103/PhysRev.130.1605"
}

@article{LukyanovZamolodchikov1997,
    author        = "Lukyanov, Sergei L. and Zamolodchikov, Alexander B.",
    title         = "{Exact expectation values of local fields in the quantum sine-Gordon model}",
    journal       = "Nucl. Phys. B",
    volume        = "493",
    pages         = "571--587",
    year          = "1997",
    doi           = "10.1016/S0550-3213(97)00123-5",
    eprint        = "hep-th/9611238",
    archivePrefix = "arXiv",
    primaryClass  = "hep-th"
}

@article{Maillet1986,
    author  = "Maillet, J. M.",
    title   = "{New integrable canonical structures in two-dimensional models}",
    journal = "Nucl. Phys. B",
    volume  = "269",
    number  = "1",
    pages   = "54--76",
    year    = "1986",
    doi     = "10.1016/0550-3213(86)90365-2"
}

@article{PhysRevD.19.3660,
    author    = "Thacker, H. B. and Wilkinson, David",
    title     = "{Inverse scattering transform as an operator method in quantum field theory}",
    journal   = "Phys. Rev. D",
    volume    = "19",
    number    = "12",
    pages     = "3660--3665",
    year      = "1979",
    publisher = "American Physical Society",
    doi       = "10.1103/PhysRevD.19.3660"
}

@article{PhysRevLett.19.1095,
    author    = "Gardner, Clifford S. and Greene, John M. and Kruskal, Martin D. and Miura, Robert M.",
    title     = "{Method for Solving the Korteweg-deVries Equation}",
    journal   = "Phys. Rev. Lett.",
    volume    = "19",
    number    = "19",
    pages     = "1095--1097",
    year      = "1967",
    publisher = "American Physical Society",
    doi       = "10.1103/PhysRevLett.19.1095"
}

@article{Polyakov1975,
    author  = "Polyakov, A. M.",
    title   = "{Interaction of Goldstone particles in two dimensions. Applications to ferromagnets and massive Yang--Mills fields}",
    journal = "Phys. Lett. B",
    volume  = "59",
    number  = "1",
    pages   = "79--81",
    year    = "1975",
    doi     = "10.1016/0370-2693(75)90161-6"
}

@article{Sklyanin:1979pfu,
    author       = "Sklyanin, E. K. and Takhtadzhyan, L. A. and Faddeev, L. D.",
    title        = "{Quantum inverse problem method. I}",
    reportNumber = "LOMI-P-1-79",
    doi          = "10.1007/BF01018718",
    journal      = "Theor. Math. Phys.",
    volume       = "40",
    number       = "2",
    pages        = "688--706",
    year         = "1979"
}

@book{Smirnov:1992vz,
    author    = "Smirnov, F. A.",
    title     = "{Form Factors in Completely Integrable Models of Quantum Field Theory}",
    publisher = "World Scientific",
    address   = "Singapore",
    series    = "Advanced Series in Mathematical Physics",
    volume    = "14",
    isbn      = "978-981-02-0244-6",
    doi       = "10.1142/1115",
    year      = "1992"
}

@article{Thacker:1980ei,
    author       = "Thacker, H. B.",
    title        = "{Exact Integrability in Quantum Field Theory and Statistical Systems}",
    reportNumber = "FERMILAB-PUB-80-038-T",
    doi          = "10.1103/RevModPhys.53.253",
    journal      = "Rev. Mod. Phys.",
    volume       = "53",
    pages        = "253",
    year         = "1981"
}

@article{Wadati:1985cc,
    author       = "Wadati, Miki and Kuniba, Atsuo and Konishi, Tetsuro",
    title        = "{The Quantum Nonlinear Schr{\"o}dinger Model: Gelfand--Levitan Equation and Classical Soliton}",
    reportNumber = "Print-85-0028 (TOKYO)",
    doi          = "10.1143/JPSJ.54.1710",
    journal      = "J. Phys. Soc. Jap.",
    volume       = "54",
    pages        = "1710",
    year         = "1985"
}

@article{WilsonKogut1974,
    author  = "Wilson, Kenneth G. and Kogut, John B.",
    title   = "{The renormalization group and the $\epsilon$ expansion}",
    journal = "Phys. Rep.",
    volume  = "12",
    number  = "2",
    pages   = "75--199",
    year    = "1974",
    doi     = "10.1016/0370-1573(74)90023-4"
}

@article{Zamolodchikov:1977py,
    author  = "Zamolodchikov, A. B.",
    title   = "{Exact Two-Particle S-Matrix of Quantum Sine-Gordon Solitons}",
    journal = "JETP Lett.",
    volume  = "25",
    pages   = "468--471",
    year    = "1977",
    note    = "[Pisma Zh. Eksp. Teor. Fiz. {\bf 25}, 499 (1977)]"
}

@incollection{Zamolodchikov:1989ASPM,
    author    = "Zamolodchikov, A. B.",
    title     = "{Integrable field theory from conformal field theory}",
    booktitle = "Integrable Systems in Quantum Field Theory and Statistical Mechanics",
    series    = "Adv. Stud. Pure Math.",
    volume    = "19",
    pages     = "641--674",
    publisher = "Mathematical Society of Japan",
    year      = "1989",
    doi       = "10.2969/aspm/01910641"
}

@article{ZamolodchikovAl1995,
    author  = "Zamolodchikov, Alexei B.",
    title   = "{Mass scale in the sine-Gordon model and its reductions}",
    journal = "Int. J. Mod. Phys. A",
    volume  = "10",
    number  = "8",
    pages   = "1125--1150",
    year    = "1995",
    doi     = "10.1142/S0217751X9500053X"
}

\end{document}